\documentclass[physrev,lengthcheck,superscriptaddress,nofootinbib,floatfix]{revtex4-2}
\usepackage[colorlinks,allcolors=BrickRed,linktocpage,unicode]{hyperref}
\usepackage[dvipsnames]{xcolor}
\usepackage{amsmath,amssymb,amsfonts,physics}
\usepackage{graphicx}
\usepackage{appendix}
\pdfoutput=1

\newcommand{\eq}[1]{\begin{align}#1\end{align}}
\newcommand{\bs}{\boldsymbol}
\newcommand{\mr}{\mathrm}
\newcommand{\UU}{\mathcal{U}}
\newcommand{\id}{\hat{\delta}}

\begin{document}

\title{Gas-liquid phase separation at zero temperature: mechanical interpretation and implications for gelation}

\author{Masanari Shimada}

\email{masanari.shimada@ryerson.ca}

\affiliation{Graduate School of Arts and Sciences, The University of Tokyo, Tokyo 153-8902, Japan}
\affiliation{Department of Physics, Toronto Metropolitan University, M5B 2K3, Toronto, Canada}

\author{Norihiro Oyama}

\affiliation{Graduate School of Arts and Sciences, The University of Tokyo, Tokyo 153-8902, Japan}
\affiliation{Mathematics for Advanced Materials-OIL, AIST, Sendai 980-8577, Japan}

\date{\today}

\begin{abstract}
The relationship between glasses and gels has been intensely debated for decades; however, the transition between these two phases remains elusive.
To investigate a gel formation process in the zero-temperature limit and its relation to the glass phase, we conducted numerical experiments on athermal quasistatic decompression.
During decompression, the system experiences a cavitation event similar to phase separation and this is a gelation process at zero temperature.
A normal mode analysis revealed that the phase separation is signaled by the vanishing of the lowest eigenenergy, similar to plastic events of glasses under shear. 
One primary difference from the shear-induced plasticity is that the vanishing mode experiences a qualitative change in its spatial energy distribution at the phase separation point. 
These findings enable us to define the glass-gel phase boundary based on mechanics.
\end{abstract}

\maketitle

\section{Introduction}

Phase separation is a key to the formation of heterogeneous structures in nature.
In particular, if there is a strong viscoelastic asymmetry between the two coexisting phases, the system exhibits various patterns during the process of phase separation~\cite{Tanaka2000Viscoelastic}.
This phenomenon is called viscoelastic phase separation and is attributed to many pattern formation phenomena such as the formation of membrane filters or plastic foams~\cite{Tanaka2000Viscoelastic}.

The viscoelastic phase separation also plays an essential role in the formation of the network-like structure of a physical gel~\cite{Zaccarelli2007Colloidal,Royall2018Vitrification,Lu2008Gelation,Zaccarelli2008Gelation,Testard2011Influence,Testard2014Intermittent,Chaudhuri2016Phase,Chaudhuri2016Structural}.
Unlike chemical gels, in which networks are formed by chemical reaction such as crosslinking, physical gels are characterized by transient networks formed by intermolecular forces.
When a liquid state is quenched deeply into the gas-liquid coexisting region and the temperature is lower than the glass transition temperature (see also Fig.~\ref{fig:phase}), the phase separation process is dynamically arrested by the slow dense phase.
Due to this slow dynamics, the phase separation cannot be completed within the experimental time period.
The resulting non-equilibrium state is interpreted as a gel.

Thus, gels are closely related to glasses~\cite{Berthier2011Theoretical,Zaccarelli2007Colloidal}, which are almost homogeneous down to the scale of their constituent particles.
However, it is difficult to clearly distinguish these two disordered states near the boundary between the two phases because low-density glasses and high-density gels exhibit almost identical structures~\cite{Zaccarelli2007Colloidal,Royall2018Vitrification}.
In other words, one cannot determine precisely the line that separates the liquid and coexisting phases in the low-temperature region based only on structural data.

Numerical simulations in the zero-temperature limit are a useful tool for investigating such low-temperature phenomena.
In this extreme limit, systems are located at local minima of the potential energy, which are called inherent structures.
In the case of glasses, for example, a normal mode analysis of inherent structures revealed that intermittent plastic events under an external shear are induced by the destabilization of spatially localized eigenmodes~\cite{Maloney2006Amorphous,Karmakar2010Statistical,Manning2011Vibrational}.
These modes are unique to glasses in the lowest-frequency region and are called \emph{quasi}-localized modes (QLMs) because of their slowly decaying tails~\cite{Lerner2016Statistics}.

In contrast to a homogeneous glass state, several studies revealed that an inherent structure experiences a zero-temperature phase separation into a dense glass phase and cavities at a certain density~\cite{Corti1997Constraints, Sastry1997Statistical,Sastry2000Liquid,Altabet2016cavitation,Altabet2018Cavitation,Gish2020Does}.
This phenomenon and the corresponding density are called the Sastry transition and Sastry density $\rho_S$, respectively~\cite{Altabet2016cavitation,Altabet2018Cavitation,Gish2020Does}.
The Sastry transition shares qualitative similarities with the conventional first-order phase transition~\cite{Altabet2016cavitation} despite the absence of thermal fluctuations; the pressure-density curve exhibits a loop as will be shown in Fig.~\ref{fig:press_a}.
The density at which this loop reaches the minimum value is the conventional definition of the Sastry density $\rho_S$.
The Sastry density can be regarded as the glass-gel phase boundary at zero temperature, and it seems promising to study the Sastry transition as a first step towards understanding the complicated process of the viscoelastic phase separation at finite temperatures.
However, the mechanism of the Sastry transition is still quite elusive because the first-order nature of the Sastry transition induces strong finite size effects and hysteretic behavior, which prevent a precise measurement of the transition point $\rho_S$~\cite{Altabet2016cavitation,Altabet2018Cavitation}.

Here, we propose a purely mechanical interpretation of the Sastry transition.
We performed molecular dynamics simulations of glasses and investigated inherent structures under quasistatic decompression at zero temperature, called athermal quasistatic (AQS)~\cite{Maeda1978Computer} decompression.
A normal mode analysis of inherent structures revealed that the Sastry transition is induced by the destabilization of the lowest-frequency QLM.
This process evolves with the same functional form as plastic events under shear, indicating that the instability is induced by a saddle-node bifurcation~\cite{Maloney2006Amorphous,Tanguy2010Vibrational,Manning2011Vibrational}.
Furthermore, we identified that the spatial energy distribution of the lowest-frequency QLM changes qualitatively near the Sastry density.
This qualitative change provides an intuitive explanation of the Sastry transition and a firm distinction between a glass and a gel on the basis of recent theoretical developments~\cite{Mueller2015Marginal}.

This paper is organized as follows.
In Sec.~\ref{sec:methods}, we introduce the model and numerical methods used in this study.
In Sec.~\ref{sec:two}, we review basic facts about the Sastry transition and introduce two protocols to reach the Sastry density adopted in this study.
In Sec.~\ref{sec:Sastry}, we investigate the lowest-frequency eigenmodes of inherent structures near the Sastry density and show that these modes are destabilized at the Sastry density.
In Sec.~\ref{sec:energetics}, we investigate the spatial energy distribution of the lowest-frequency eigenmodes and discuss its density dependence.
In Sec.~\ref{sec:discussion}, we discuss the relation between the results in Sec.~\ref{sec:energetics} and the thermodynamic limit.
In Sec.~\ref{sec:summary}, we conclude the paper with a summary.
In Appendix~\ref{sec:structural}, we confirm that our system does not crystallize.
In Appendix~\ref{sec:harmonic}, we provide the fundamental aspects of the normal mode analysis in detail.
In Appendix~\ref{sec:correlations}, we give additional data to supplement Fig.~\ref{fig:snap_all}.
In Appendix~\ref{sec:energy}, we show the results of different system sizes corresponding to Fig.~\ref{fig:energy}.

\section{Methods}\label{sec:methods}

We used three-dimensional (3D) monodisperse particles with mass $m$ interacting via a Lennard-Jones (LJ) potential, $\phi_0(r) = 4\epsilon \left[ (\sigma/r)^6 - (\sigma/r)^{12} \right]$, where $\epsilon$ and $\sigma$ are the characteristic energy and length scales, respectively.
Below, length, mass, and time are reported in units of $\sigma$, $m$, and $\sqrt{m\sigma^2/\epsilon}$, respectively.
We truncated $\phi_0(r)$ at $r_c=2.5$ and shifted it so that the resulting potential and its first derivative continuously tend to zero at $r=r_c$,
\eq{
\phi(r) = 
    \begin{cases}
        \phi_0(r) - \phi_0(r_c) - \phi'_0(r_c)(r-r_c) & (r<r_c) \\
        0 & (r>r_c)
    \end{cases}
    .
}
We performed molecular dynamics simulations with this potential.
100 equilibrium liquid configurations were generated for different system sizes, from $N=1000$ to $64000$, at a sufficiently high temperature $T=2.0$.
Starting from these normal liquids, we performed instantaneous quenches to zero temperature using the steepest descent method and prepared inherent structures.

Since we used a monodisperse system, crystallization might be a concern.
However, the fraction of microscopic crystalline structures measured by the locally averaged bond orientational order parameters~\cite{Lechner2008Accurate, Kawasaki2010Formation} is sufficiently small in our system, as discussed in Appendix~\ref{sec:structural}.
Thus we conclude that our system has no crystalline order.
The quench rate to avoid the crystallization of a monodisperse LJ system is discussed in detail in Ref.~\cite{Monaco2009Anomalous}.

We then conducted a normal mode analysis~\cite{Kittel2004Introduction} of those inherent structures.
We diagonalized the dynamical matrix, which is the second derivative of the total potential $\UU$ around an inherent structure, to obtain its eigenvalues $\lambda_\alpha$ and eigenvectors $\bs{e}_\alpha = (\bs{e}_{\alpha,1},\ldots,\bs{e}_{\alpha,N})$, where $\alpha=1,2,\ldots,3N-3$.
Note that we excluded three modes corresponding to the global translations.
The eigenvectors were normalized: $|\bs{e}_{\alpha}|=1$.
The eigenfrequency is given by $\omega_\alpha=\sqrt{\lambda_\alpha}$.
In Appendix~\ref{sec:harmonic}, we provide the fundamental aspects of the normal mode analysis in detail.

\section{Results}

\subsection{Two approaches to Sastry transition}\label{sec:two}

\begin{figure}
    \centering
    \includegraphics[width=\linewidth]{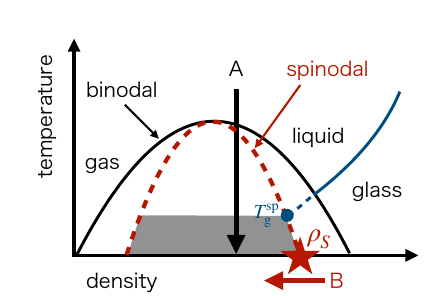}
    \caption{
    Schematic phase diagram of a particulate system with attractive interaction.
    The binodal, spinodal, and glass transition lines are displayed on the temperature-density plane.
    $T_{\mr{g}}^{\mr{sp}}$ indicates the intersection of the extrapolated glass transition line and the spinodal line.
    Arrow A indicates a quench from a normal liquid, and arrow B indicates an AQS decompression.
    The star symbol marks the Sastry density $\rho_S$.
    }
    \label{fig:phase}
\end{figure}

We present a schematic phase diagram of a typical attractive system~\cite{Zaccarelli2007Colloidal} in Fig.~\ref{fig:phase}.
The binodal, spinodal, and glass transition lines are displayed on the temperature-density plane.
The temperature $T_{\mr{g}}^{\mr{sp}}$ is the intersection of the extrapolated glass transition line and the spinodal line.
Before discussing the detail of this figure, we make the following two remarks.
First, this is a non-equilibrium phase diagram.
In the equilibrium phase diagram, the glass transition and spinodal lines are not well-defined~\cite{Binder1987Theory,Binder2012Beyond}.
These lines are usually defined based on the dynamics.
The glass transition temperature is the one at which the relaxation time or the viscosity exceeds a threshold value, e.g., an experimentally accessible upper limit.
Also, if the system is quenched to a temperature below the spinodal line, phase separation proceeds by spinodal decomposition\footnote{
This is a standard definition of the spinodal at finite temperatures.
Also if we focus on the zero-temperature limit as in this study, the spinodal is a rigorous notion~\cite{Nandi2016Spinodals}.}.
Second, the precise locations of the binodal and spinodal lines at these low temperatures still remain to be determined~\cite{Testard2014Intermittent,Lu2008Gelation,Zaccarelli2008Gelation} and Fig.~\ref{fig:phase} presents a very simplified diagram.

We now describe two processes that were used in this study to reach the gray shaded region below $T_{\mr{g}}^{\mr{sp}}$, where the phase separation dynamics are arrested by the slow glass phase.
The first one, indicated by arrow A in Fig.~\ref{fig:phase}, is a quench from a normal liquid at a fixed density.
We particularly consider a quench to zero temperature in this study.
The other one, indicated by arrow B, is an AQS decompression from a glass.
Using these two processes, the Sastry transition is observed when the system crosses the zero-temperature limit of the spinodal line, the Sastry density $\rho_S$, marked by the star symbol~\cite{Altabet2018Cavitation,Altabet2016cavitation,Altabet2018Cavitation,Gish2020Does}.
In this paper, we call the former process A and the latter process B.

\begin{figure}
    \centering
    \includegraphics{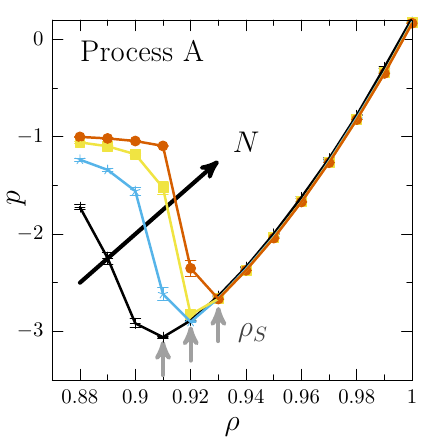}
    \caption{
    Pressure $p$ versus density $\rho$ curves of inherent structures for $N=1000$, 4000, 16000, and 64000 obtained by the instantaneous quench, or process A.
    The $N$-dependent Sastry densities $\rho_S$ are indicated by three up arrows.
    Note that the Sastry density for $N=4000$ is the same as that for $N=16000$ with this resolution of density.
    }
    \label{fig:press_a}
\end{figure}

With the phase diagram in Fig.~\ref{fig:phase} in mind, we next recapitulate established knowledge of the Sastry transition~\cite{Sastry1997Statistical,Sastry2000Liquid,Altabet2016cavitation,Altabet2018Cavitation} using our data.
We present pressure $p$ versus density $\rho$ curves in Fig.~\ref{fig:press_a}.
This figure shows the data for $N=1000$, 4000, 16000, 64000 obtained by the instantaneous quench, or process A.
We can see that the pressure monotonically decreases from $\rho=1.0$ to $\rho\sim 0.93$ independently of the system size $N$.
Note that with periodic boundary conditions homogeneous states can be stable even though the pressure is negative~\cite{Sastry2000Liquid}.
However, for $\rho\lesssim0.93$, the pressure reaches its minimum, whose location strongly depends on $N$, and subsequently increases.
The system has cavities in this regime~\cite{Sastry1997Statistical,Sastry2000Liquid,Altabet2016cavitation,Altabet2018Cavitation}.
Therefore, the density corresponding to the minimum pressure can be interpreted as the zero-temperature phase separation point, which is the conventional definition of the Sastry density $\rho_S$~\cite{Altabet2016cavitation,Altabet2018Cavitation,Gish2020Does}.
The Sastry density for each $N$ is indicated by an up arrow in Fig.~\ref{fig:press_a}.
Note that the Sastry density for $N=4000$ is the same as that for $N=16000$ with this resolution of density.

\begin{figure}
    \centering
    \includegraphics{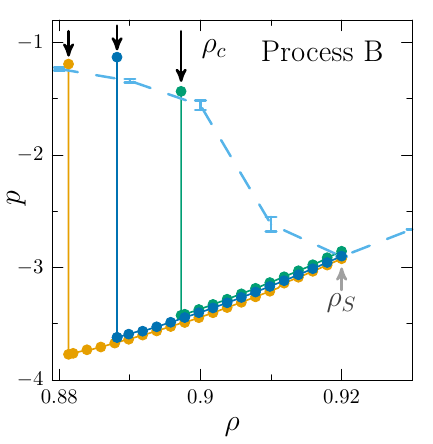}
    \caption{
    Pressure $p$ versus density $\rho$ curves of inherent structures for $N=4000$ obtained by the AQS decompression, or process B.
    Trajectories of three samples out of 100 are shown by filled circles and for each trajectory, the cavitation point $\rho_c$ is indicated by a down arrow.
    For comparison, the corresponding data obtained by the process A is shown by the dashed line, see also Fig.~\ref{fig:press_a}.
    }
    \label{fig:press_b}
\end{figure}

\begin{table}
    \renewcommand{\arraystretch}{1.2}
    \setlength\tabcolsep{0.5em}
    \centering
    \caption{
    Definitions of three characteristic densities related to the Sastry transition.
    }
    \begin{tabular}{ccc} \hline \hline
        Symbol      & Description & Figure \\ \hline
        $\rho_S$    & Sastry density & Fig.~\ref{fig:press_a} \\
        $\rho_c$    & Cavitation point & Fig.~\ref{fig:press_b} \\
        $\rho_\ast$ & Destabilization of parallel motions & Fig.~\ref{fig:energy} \\ \hline \hline
    \end{tabular}
    \label{tab:densities}
\end{table}

For the AQS decompression, or process B, we used 100 configurations of $N=4000$ obtained by the process A at $\rho=0.92$ as initial states.
We decompressed these configurations by repeating a very small reduction of the density, by a value of $\Delta \rho$, followed by the minimization of the total energy of the system.
We set the initial relative density decrement to $\Delta\rho/\rho\simeq10^{-4}$ and detected a cavitation event by reducing the relative density decrement to $10^{-8}$, which follows the backtracking procedure~\cite{Lerner2009Locality,Karmakar2010Statisticala}.
Figure~\ref{fig:press_b} shows the results of this procedure.
We present trajectories of three samples out of 100 by filled circles and the corresponding data obtained by the process A is shown by the dashed line.
The pressure continues to decrease in this process even below the Sastry density $\rho_S$ and at a certain density $\rho_c \ll \rho_S$, it jumps to a value comparable to the ones obtained by the process A.
At this density $\rho_c$, the system forms a cavity.
Practically, we defined the cavitation point $\rho_c$ as the density at which the pressure increases by more than 50 percent during the process B.
Since the cavitation point $\rho_c$ shows large sample-to-sample fluctuations, it will be interesting future work to investigate their statistics.
In Tab.~\ref{tab:densities}, we summarized three characteristic densities related to the Sastry transition: $\rho_S$, $\rho_c$, and $\rho_\ast$. 
The third one, $\rho_\ast$, will be defined later in Sec.~\ref{sec:energetics}.

\subsection{Sastry transition as a mechanical instability}\label{sec:Sastry}

\begin{figure}[t]
    \centering
    \includegraphics{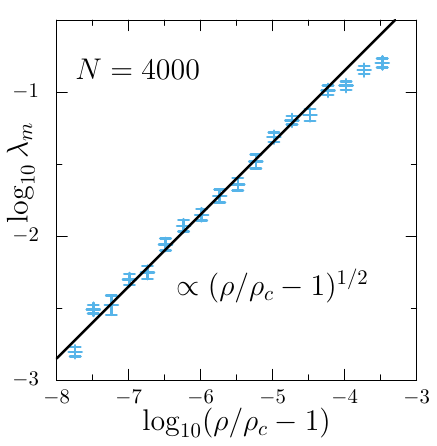}
    \caption{
    The sample-averaged smallest eigenvalue $\lambda_{m}$ as a function of the relative density $\rho/\rho_c-1$.
    The solid line indicates a power law $\lambda_m\propto(\rho/\rho_c-1)^{1/2}$.
    }
    \label{fig:eigen}
\end{figure}

The density dependence of the lowest-frequency eigenmode upon decompression provides an important insight into the Sastry transition.
As mentioned in the introduction, the lowest-frequency modes of glasses are quasi-localized~\cite{Lerner2016Statistics}.
To investigate the lowest-frequency eigenmode, we used 100 trajectories of the process B mentioned in the last paragraph of the previous section and Fig.~\ref{fig:press_b}.
We computed the smallest eigenvalues $\lambda_{m}$ of these configurations during the AQS decompression, or process B.
In Fig.~\ref{fig:eigen}, we depict the dependence of $\lambda_{m}$ on the relative density $\rho/\rho_c-1$, where $\rho_c$ is the cavitation point which was defined in the previous section, see Tab.~\ref{tab:densities}.
We obtained $\lambda_m$ versus $\rho/\rho_c-1$ data for each sample and averaged all the data for 100 samples to plot Fig.~\ref{fig:eigen}.
The data clearly obey a power law $\lambda_{m}\propto(\rho/\rho_c-1)^{1/2}$, which is indicated by the solid line, and $\lambda_{m}$ vanishes at $\rho=\rho_c$.
This power law can be derived from bifurcation theory and is well-established in the case of shear-induced plasticity~\cite{Maloney2006Amorphous}.
These results indicate that the Sastry transition can be interpreted as an event induced by a global mechanical instability similar to a plastic event under shear.

\begin{figure}
    \centering
    \includegraphics{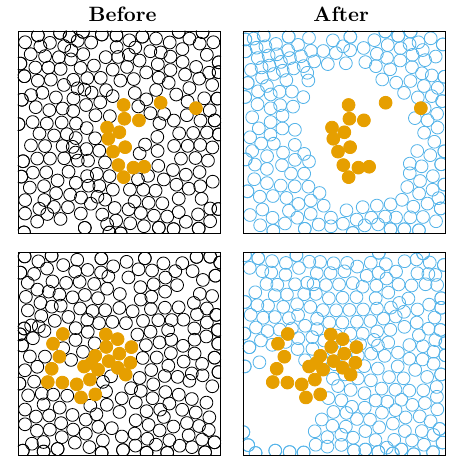}
    \caption{
    Unfilled circles: 2D slices of two sets of $4000$-particle configurations immediately before (left) and after (right) the cavitation. 
    Filled circles: the $\lceil N P_r\rceil$ particles with the largest components in the lowest-frequency QLMs before the cavitation.
    The difference between the configurations in the left and right columns is of order $10^{-8}$ in relative density.
    }
    \label{fig:snap_all}
\end{figure}

Since we showed that the lowest-frequency modes are destabilized at the cavitation points, we now investigate what happens in real-space structures during the cavitation events.
In Fig.~\ref{fig:snap_all}, we show 2D slices of configurations immediately before and after the cavitation using unfilled circles (see Appendix~\ref{sec:correlations} for four more samples).
In addition, the lowest-frequency QLM in each sample before the cavitation is shown using filled circles.
To visualize these QLMs, contributing particles were determined using the participation ratio, $P_r = (\sum_i \bs{v}_i^2)^2/(N\sum_i \bs{v}_i^4)$, where $\bs{v}$ is a $3N$-dimensional vector. 
This is a measure of the localization of vibrational modes~\cite{Mazzacurati1996Low}.
When all particles vibrate equally, $P_r=1$ and when only one particle vibrates, $P_r=1/N$.
In Fig.~\ref{fig:snap_all}, we show $\lceil N P_r\rceil$ particles that have the largest components in the QLMs.
Figure~\ref{fig:snap_all} presents clear, though not perfect, correlations between the positions of the cavities and the QLMs.

\begin{figure}[t]
    \centering
    \includegraphics[width=0.45\textwidth]{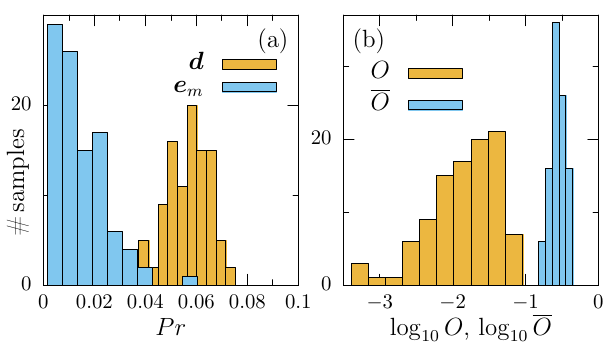}
    \caption{
    (a) Histograms of the participation ratio of the displacement during the cavitation $\bs{d}$ and the lowest-frequency eigenvector $\bs{e}_{m}$.
    (b) Histograms of the overlap $O$ between $\bs{d}$ and $\bs{e}_m$ and the reduced overlap $\overline{O}$.
    Results are drawn from the $N=4000$ system.
    Statistics of 100 samples are shown.
    }
    \label{fig:overlappdf}
\end{figure}

To quantify the correlations between the lowest-frequency QLM and the cavitation event, we measured the participation ratio of the lowest-frequency QLM $\bs{e}_{m}$ and that of the displacement during the cavitation $\bs{d}$.
Figure~\ref{fig:overlappdf}(a) presents their histograms.
We can observe that a larger number of particles are involved in $\bs{d}$ compared with $\bs{e}_{m}$.
Even though the QLMs trigger the cavitation, the induced displacements do not stop until the system finds another inherent structure; hence, they involve highly anharmonic motions.
For this reason, the correspondence between the QLMs and the cavities is not perfect as observed in Fig.~\ref{fig:snap_all}.

However, if we appropriately define an overlap between $\bs{e}_{m}$ and $\bs{d}$, we can characterize the correlations.
The conventional overlap is usually defined as the simple inner product between two vectors $O = \bs{e}_m\cdot\bs{d}/|\bs{d}|$.
In contrast, we introduce a reduced vector $\bs{\overline{v}} = (|\bs{v}_{1}|,|\bs{v}_{2}|,\ldots,|\bs{v}_{N}|)$, which retains the amplitudes only.
Based on the reduced vectors, we define a reduced overlap as $\overline{O} = \bs{\overline{e}}_1\cdot\bs{\overline{d}}/|\bs{\overline{d}}|$.
In Fig.~\ref{fig:overlappdf}(b), we present the histograms of the normal and reduced overlaps.
We can observe that the latter $\sim0.3$ is significantly larger than the former $\sim10^{-2}$.
Thus, the amplitudes between $\bs{e}_m$ and $\bs{d}$ are correlated, even though their directions are not.
The eigenvector $\bs{e}_m$ tells us which particles are mobile during the cavitation event, but it has little information about the directions of those particle motions due to the strong anharmonicity of the event.
This is the essence of the clear, but not perfect correlations observed in Fig.~\ref{fig:snap_all}.

\begin{figure*}
    \centering
    \includegraphics[width=0.9\textwidth]{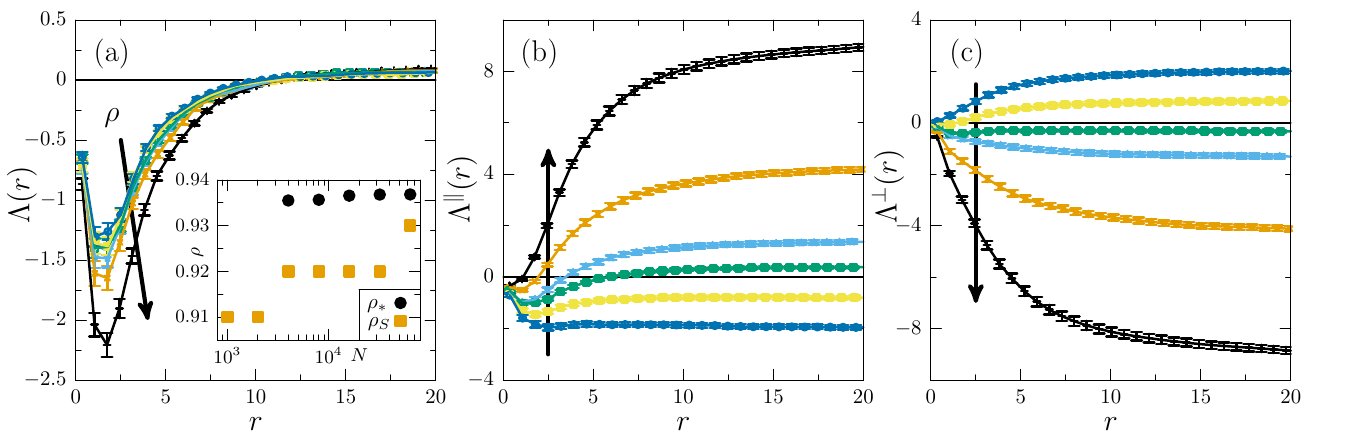}
    \caption{
    (a) Total energy profiles $\Lambda(r)$ averaged over the lowest-frequency QLMs in configurations of $N=64000$.
    The estimated values of $\rho_\ast$ and $\rho_S$ are shown in the inset figure.
    (b) Energy profiles $\Lambda^\parallel(r)$ of parallel components.
    (c) Energy profiles $\Lambda^\perp(r)$ of perpendicular components.
    The data for $\rho=0.92$, 0.93, 0.94, 0.95, 0.97, and 1.0 are shown.
    In all panels, the density $\rho$ increases as indicated by the arrows.
    }
    \label{fig:energy}
\end{figure*}

\subsection{Energetics}\label{sec:energetics}

We have thus far shown that the lowest-frequency QLM initiates a catastrophic event, which is similar to a plastic event under shear~\cite{Maloney2006Amorphous,Tanguy2010Vibrational,Manning2011Vibrational}.
However, we have also observed the formation of a cavity~\cite{Altabet2018Cavitation} as opposed to the shear-induced plasticity.
Here, we demonstrate that this qualitative difference is explained by the spatial distribution of the vibrational energy of the lowest-frequency QLM~\cite{Shimada2018Spatial, Shimada2021Spatial}.
For this purpose, we introduce the local vibrational energy~\cite{Shimada2018Spatial,Wyart2005Effects} of a particle $i$ in a mode $\bs{e}$ as\footnote{For better readability, we omit the mode index $\alpha$ hereafter.} 
\eq{
    \delta E_{i} & = \frac{1}{2} \sum_{j=1}^N \left[ \phi''_{ij}(r_{ij}) ( \bs{n}_{ij}\cdot\bs{e}_{ij} )^2 + \frac{\phi'_{ij}(r_{ij})}{r_{ij}}( \bs{e}_{ij}^\perp )^2\right] \notag \\
    & \equiv \delta E_i^\parallel + \delta E_i^\perp, \label{local}
}
where $\bs{n}_{ij} = \bs{r}_{ij}/r_{ij}$ is the unit vector pointing from a particle $j$ to $i$, $\bs{e}_{ij}=\bs{e}_{i}-\bs{e}_{j}$, and $(\bs{e}_{ij}^\perp)^2 = (\bs{e}_{ij})^2-(\bs{n}_{ij}\cdot\bs{e}_{ij})^2$.
In the second line of Eq.~\eqref{local}, we decomposed the local energy into the parallel part $\delta E_i^\parallel = \frac{1}{2} \sum_j  \phi''_{ij}(r_{ij}) ( \bs{n}_{ij}\cdot\bs{e}_{ij} )^2$ and perpendicular part $\delta E_i^\perp = \frac{1}{2} \sum_j \left[ \phi'_{ij}(r_{ij})/r_{ij} \right] ( \bs{e}_{ij}^\perp )^2$.
In purely repulsive systems, we have $\delta E_i^\parallel>0$ and $\delta E_i^\perp<0$, but the opposite signs are possible in models with attractive interactions like ours.
The parallel part corresponds to the ordinary elastic energy of a relaxed spring while the perpendicular part corresponds to the residual stress responsible for buckling-like motion~\cite{Wyart2005Effects}.
We call the particle with the most negative $\delta E_i$ the center of the mode and denote its position by $\bs{r}^c$.
This center particle typically has the largest amplitude in the QLM~\cite{Shimada2018Spatial, Shimada2021Spatial}.
The energy profile~\cite{Shimada2018Spatial, Shimada2021Spatial} is then defined as
\eq{
    \Lambda(r) & = \sum_{i = 1}^N \theta(r-|\bs{r}_i-\bs{r}^c|) \delta E_{i}  = \int_{|\bs{x}| < r} d\bs{x} \delta E (\bs{x}), \label{energy profile}
}
where $\theta(x)$ is the Heaviside step function (see also Appendix~\ref{sec:harmonic}).
In the rightmost expression, we rewrote the function using a spatial integral of the local energy density $\delta E(\bs{r}) = \sum_i \delta E_i \delta[ \bs{r} - (\bs{r}_i-\bs{r}^c) ]$.
Thus, $\Lambda(r)$ is the vibrational energy that the QLM would have if the system was cut at a distance $r$ from the center $\bs{r}^c$.
Note that $\Lambda(r)$ converges to the eigenvalue of the mode $\bs{e}$ as $r\to\infty$.
We also define the parallel and perpendicular energy profiles, $\Lambda^\parallel(r)$ and $\Lambda^\perp(r)$, by replacing $\delta E_i$ with $\delta E^\parallel_i$ and $\delta E^\perp_i$ in Eq.~\eqref{energy profile}, respectively.

Figure~\ref{fig:energy} shows the (a) total, (b) parallel, and (c) perpendicular energy profiles for the systems of $N=64000$ at different densities from $\rho=0.92$ to 1.0.
The presented data are averages over 100 samples, which were obtained using the instantaneous quench or process A\footnote{To average the data for $\rho=0.92$, we excluded 33 configurations that were already cavitated. As shown in Fig.~\ref{fig:press_a}, this density is already lower than the Sastry density for $N=64000$.}.
Figure~\ref{fig:energy}(a) shows that the total energy profile starts from a negative value, reaches its minimum, transitions to a positive value, and finally converges to the average eigenvalue.
Qualitatively, the same behavior is observed in the QLMs of repulsive systems and the length at which $\Lambda(r)$ achieves a minimum has been used as the definition for the core size of the QLMs~\cite{Shimada2018Spatial}.
Importantly, the total energy profile hardly depends on the density, and the core size is always $r\sim2$ throughout the whole density range.
This density (in)dependence is markedly different from that of repulsive systems close to unjamming, in which the core size diverges at the unjamming transition~\cite{Shimada2018Spatial}.

However, $\Lambda^\parallel(r)$ and $\Lambda^\perp(r)$ exhibit strong density dependences.
At $\rho=1.0 \gg \rho_S$, the former is positive, except near the origin, whereas the latter is completely negative.
Note that if either of $\Lambda^\parallel(r\to\infty)$ or $\Lambda^\perp(r\to\infty)$ is negative, the other is always positive because the total energy profile $\Lambda(r\to\infty)$ is positive due to stability.
These functional forms are similar to those in purely repulsive systems. 
This indicates that repulsive forces are dominant in determining the energetics of the QLMs and that the attractive forces can be treated as a perturbation in this dense regime.
However, as the density decreases, the values of $\Lambda^\parallel(r)$ diminish and become negative at a density $\rho=\rho_\ast\sim0.94$.
Correspondingly, $\Lambda^\perp(r)$ becomes positive.
To estimate $\rho_\ast$, we fitted a linear function to the final values of $\Lambda^\parallel(r)$ at $\rho=0.92$, 0.93, 0.94, and 0.95.
As shown in the inset of Fig.~\ref{fig:energy}(a), the estimated value of $\rho_\ast$ hardly depends on the system size $N$ while the Sastry density $\rho_S$, which is defined as the minimum of the $p$-$\rho$ curve in Fig.~\ref{fig:press_a}, strongly does\footnote{The estimation of $\rho_S$ will not qualitatively improve, even if the number of density points is increased.
As an example, Fig.~\ref{fig:press_a} shows that the pressure at $\rho=0.93$ is already larger than that at $\rho=0.92$ for $N=1000$}.
In Appendix~\ref{sec:energy}, we show that these results are qualitatively independent of the system size by directly comparing results for different $N$.
In the next section, we discuss that $\rho_\ast$ is expected to be the zero-temperature phase separation point in the thermodynamic limit.
Finally, we again emphasize that this qualitative change in the energetics of the QLMs is unique to systems with attractive interaction.

\section{Discussion}\label{sec:discussion}

\begin{table}[hb]
    \renewcommand{\arraystretch}{1.2}
    \setlength\tabcolsep{0.5em}
    \centering
    \caption{
    Number of samples for each $N$ used to compute the vibrational spectrum in Fig.~\ref{fig:dos}.
    }
    \begin{tabular}{cc} \hline \hline
        $N$   & \#samples \\ \hline
        4000, 8000  & 4000 \\
        16000       & 2000 \\
        32000       & 1000 \\
        64000       & 500 \\ \hline \hline
    \end{tabular}
    \label{tab:samples}
\end{table}

\begin{figure}[t]
    \centering
    \includegraphics[width=0.45\textwidth]{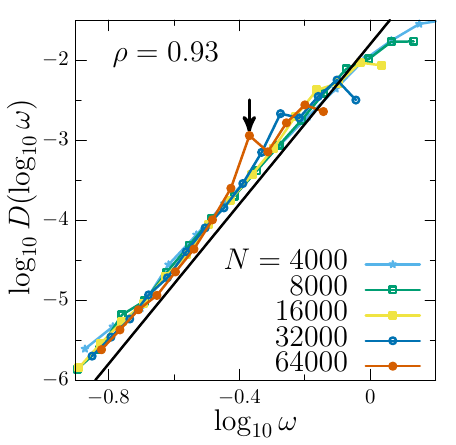}
    \caption{
    Vibrational spectrum as a function of the logarithm of the frequency, $\log_{10}\omega$, at $\rho=0.93$.
    The solid line depicts the expected behavior of the QLMs, $D_{\mr{QLM}}(\omega)\propto\omega^4$~\cite{Lerner2016Statistics,Shimada2018Anomalous}.
    The arrow indicates the peak of the lowest-frequency phonons for $N=64000$.
    For the number of samples used in this computation, see Tab.~\ref{tab:samples}.
    }
    \label{fig:dos}
\end{figure}

In this section, we first confirm that the vibrational spectrum of the QLMs follows the quartic law $D_{\mr{QLM}}(\omega)\propto\omega^4$~\cite{BaityJesi2015Soft,Lerner2016Statistics,Mizuno2017Continuum,Shimada2018Anomalous,Wang2019Low,Richard2020Universality,Das2020Robustness} even in the vicinity of the Sastry transition.
This power law is crucial for the discussion below.
Figure~\ref{fig:dos} shows the lowest-frequency tail of the vibrational spectrum $D(\omega)$ at $\rho=0.93$.
Table~\ref{tab:samples} shows the number  of samples required to obtain the data in Fig.~\ref{fig:dos}.
This figure shows that the lowest-frequency tail of the vibrational spectrum indeed obeys the quartic law.

We are now in a position to discuss implications of our results in Sec.~\ref{sec:energetics} in the thermodynamic limit.
In this limit, it is important that the density of the QLMs obeys the power law $D_{\mr{QLM}}(\omega)\propto\omega^4$ down to zero frequency~\cite{Lerner2016Statistics,Mizuno2017Continuum,Shimada2018Anomalous,Wang2019Low,Richard2020Universality,Das2020Robustness} as observed in Fig.~\ref{fig:dos}.
Such a \emph{gapless} power law indicates an abundance of arbitrarily soft modes, and a phenomenological argument suggests that systems with a gapless distribution are extremely susceptible to perturbations~\cite{Mueller2015Marginal}.
This extreme susceptibility is termed as the \emph{marginal stability} and has been studied for decades, particularly in the field of replica theory~\cite{Charbonneau2014Fractal,Berthier2019Gardner,Parisi2020Theory}.
The marginal stability is now considered as one of the fundamental properties of many disordered materials~\cite{Mueller2015Marginal}.
In fact, numerical studies suggest that an infinitesimally small shear strain can cause a plastic event due to the lowest-frequency QLM in the thermodynamic limit~\cite{Karmakar2010Statistical,Lerner2018Protocol,Shang2020Elastic,Oyama2021Unified}.
Therefore, we expect that the cavitation event occurs at the same time as $\Lambda^\parallel(r\to\infty)$ turns negative and parallel motions are globally destabilized at $\rho=\rho_\ast$ because in the thermodynamic limit this mode is susceptible to infinitesimal perturbations including decompression, which directly couples to such parallel motions and leads to instability in the density field.
In other words, $\rho_S$ and $\rho_\ast$ should coincide in the thermodynamic limit; hence, we can regard $\rho_\ast$ as the zero-temperature phase separation point.

\section{Summary and conclusion}\label{sec:summary}

In this study, we investigated the Sastry transition, which is interpreted as the zero-temperature limit of the gas-liquid phase separation.
If the density $\rho$ of an attractive system is decreased at zero temperature, the system forms a cavity at a certain density.
This cavitation process is the Sastry transition and is expected to be strongly related to the formation of a physical gel.
Using a standard LJ potential, we studied this cavitation process and found that the lowest-frequency eigenvalue $\lambda_m$ of the system vanishes at the cavitation point $\rho = \rho_c$.
The density dependence of $\lambda_m$ is a power law $\lambda_m \propto (\rho/\rho_c - 1)^{1/2}$, which is the same functional form as in the case of plastic events of glasses under shear.
In real space, the positions of the lowest-frequency eigenmode and the cavity are correlated.
To further investigate the mechanism of the cavitation, we studied the spatial energy profile of the lowest-frequency eigenmode and found that the motion parallel to particle bonds $\bs{n}_{ij}$ is globally destabilized at a density $\rho = \rho_\ast$.
Based on the notion of marginal stability, we argued that $\rho_\ast$ converges to the Sastry density $\rho_S$ in the thermodynamic limit.
Since $\rho_\ast$ is not strongly affected by the system size, this is a good estimate of the Sastry density, i.e., the zero-temperature gelation point.

As mentioned in the introduction, the zero-temperature gas-liquid phase separation investigated in this study is an extreme example of the viscoelastic phase separation.
Since our mechanical methods are not restricted to the present model, it would be interesting future work to study systems that exhibit the viscoelastic phase separation extensively.
We would be able to understand the formation of heterogeneous materials such as membranes and foams on an equal footing.

In the context of the shear-induced plasticity, the correlation between plastic events is known to self-organize into avalanches and leads to the so-called yielding criticality as the external shear reaches a threshold value, while those plastic events are localized and the criticality is absent in an isotropic unperturbed state~\cite{Karmakar2010Statistical,Oyama2021Unified}.
Since the onset of the cavitation shares the same phenomenological origin with that of the plastic events under shear, it would be meaningful to investigate similar statistics of the sample-to-sample fluctuations presented in Fig.~\ref{fig:press_b}.

\section*{Acknowledgments}

We thank A. Ikeda, H. Mizuno, T. Kawasaki, and M. Hachiya for fruitful discussions. 
MS is grateful to E. De Giuli for his useful comments.
This work was supported by JSPS KAKENHI Grant Numbers 19J20036, 20K14436, and 20J00802, and by Initiative on Promotion of Supercomputing for Young or Women Researchers, Supercomputing Division, Information Technology Center, The University of Tokyo.

\section*{Author declarations}

\subsection*{Conflict of interest}

There are no conflicts to declare.

\section*{Data availability}

The data that support the findings of this study are available from the corresponding author upon reasonable request.

\clearpage
\onecolumngrid
\appendix

\section{Structural analysis}\label{sec:structural}

\begin{figure*}[hb]
    \centering
    \includegraphics[width=0.7\textwidth]{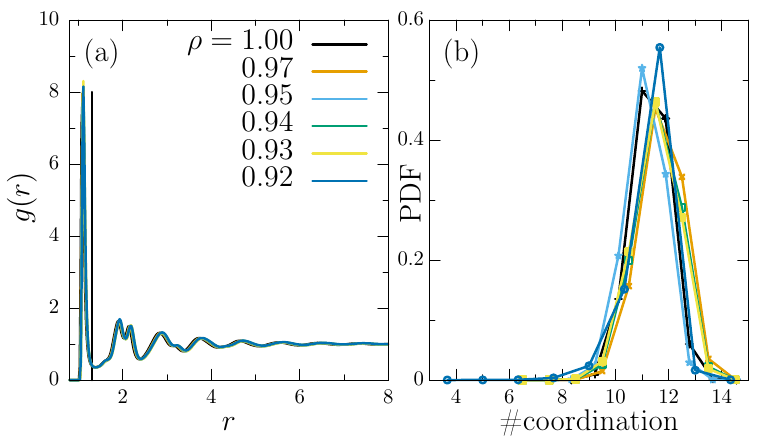}
    \caption{
    (a) Radial distribution functions for $N=64000$.
    The solid vertical line at $r_n=1.3$ indicates a cutoff to define the coordination number.
    (b) PDFs of the coordination number. 
    We counted all neighboring particles within the cutoff $r_n=$1.3.
    }
    \label{fig:app_coord}
\end{figure*}

To confirm that our system does not crystalize, we performed a structural analysis using 100 samples of $N=64000$.
Figure~\ref{fig:app_coord}(a) shows the radial distribution function $g(r)$.
It does not depend on the density and rapidly converges to one, which means that there are no long-range correlations characteristic to crystals.

\begin{figure*}
    \centering
    \includegraphics[width=0.7\textwidth]{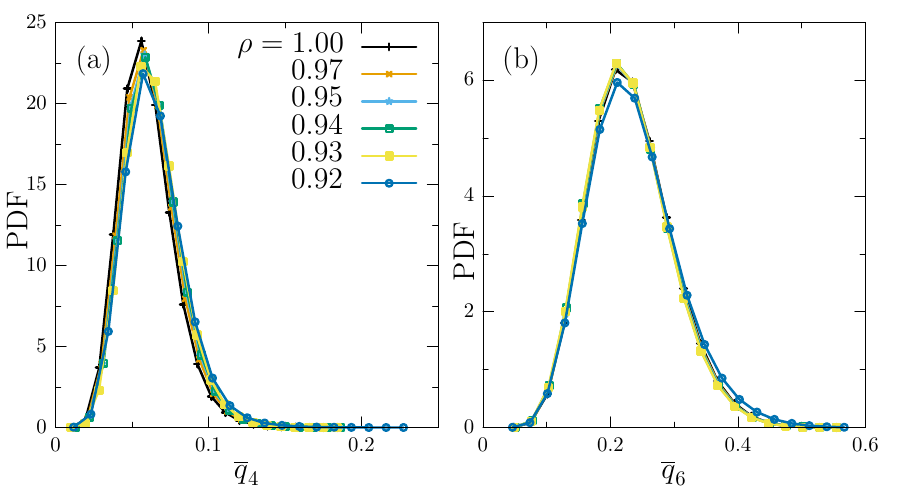}
    \caption{
    PDFs of the locally averaged bond orientational order parameters for the systems of $N=64000$.
    }
    \label{fig:app_orient}
\end{figure*}

To investigate structures at short length scales, we computed the number of neighboring particles within a cutoff $r_n=1.3$ shown in the vertical line in Fig.~\ref{fig:app_coord}(a).
Figure~\ref{fig:app_coord}(b) shows their probability distribution functions (PDFs).
Furthermore, we computed the locally averaged bond orientational order parameters $\overline{q}_n$~\cite{Lechner2008Accurate, Kawasaki2010Formation} for $n=4$ and 6 using these neighboring particles.
The PDFs of $\overline{q}_4$ and $\overline{q}_6$ shown in Fig.~\ref{fig:app_orient} indicate that both of them are small on average compared to those of crystalline structures~\cite{Lechner2008Accurate}.
However, the PDF of $\overline{q}_4$ slightly shifts to larger values when the density decreases, and that of $\overline{q}_6$ also extends to larger values at $\rho=0.92$.
This suggests that there are some, though a small fraction, crystalline structures at the microscopic scale when the density decreases.

\begin{figure*}
    \centering
    \includegraphics[width=0.7\textwidth]{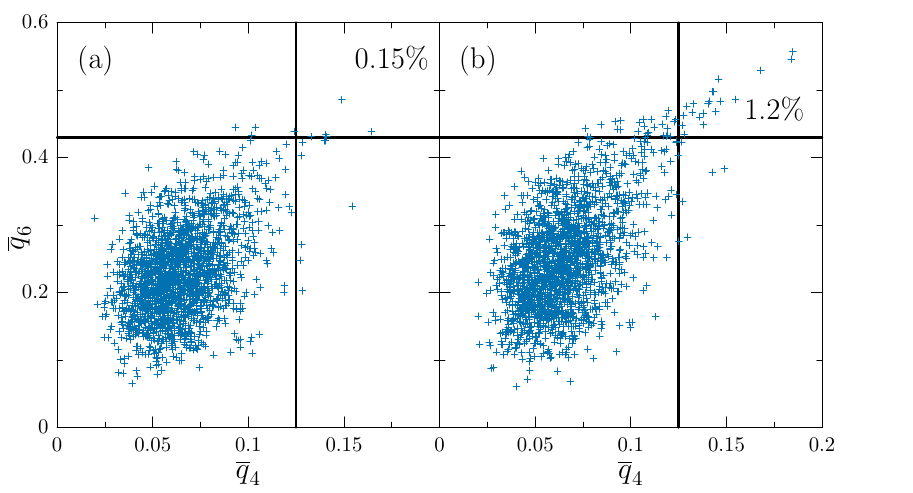}
    \caption{
    Scatter plots of the locally averaged bond orientational order parameters for (a) a homogeneous and (b) a cavitated configuration of $N=64000$ at $\rho=0.92$.
    2000 points out of 64000 from each structure were chosen randomly.
    The threshold for fcc structures, $\overline{q}_4=0.125$ and $\overline{q}_6=0.43$, is indicated by the solid lines, and the fractions of particles that exceed it are shown in percentage.
    }
    \label{fig:app_scatter}
\end{figure*}

We can precisely detect the microscopic crystalline structures by measuring the correlation between $\overline{q}_4$ and $\overline{q}_6$~\cite{Lechner2008Accurate}.
Figure~\ref{fig:app_scatter} shows scatter plots of $\overline{q}_4$ versus $\overline{q}_6$ for (a) a homogeneous and (b) a cavitated configuration of $N=64000$ at $\rho=0.92$.
2000 points out of 64000 from each structure were chosen randomly.
When a particle has large $\overline{q}_4\gtrsim0.125$ and $\overline{q}_6\gtrsim0.43$, the local structure around it is fcc~\cite{Lechner2008Accurate}.
In Fig.~\ref{fig:app_scatter}, the threshold for fcc structures is indicated by the solid lines, and the fractions of particles that exceed it are shown in percentage.
We can see that low-density configurations, including a cavitated one, have a small fraction of microscopic crystalline structures.
Thus, as long as we focus on averaged quantities, our system only has a negligible fraction of crystalline structures even at the microscopic level.

\clearpage

\section{Harmonic energy}\label{sec:harmonic}

Here we provide some fundamental aspects of the normal mode analysis~\cite{Kittel2004Introduction} and their relation to the energy profile in Eq.~\eqref{energy profile}~\cite{Shimada2018Spatial,Shimada2021Spatial}.
The total potential energy of the system is given by
\eq{
    \UU = \sum_{i>j} \phi(r_{ij}), \label{tot}
}
where $r_{ij} = | \bs{r}_i - \bs{r}_j |$ is the distance between a pair $\langle ij \rangle$.
Because the focus of this study is on inherent structures, we assume that particles are always at mechanical equilibrium, i.e., the force balance condition, $\sum_j \phi'(r_{ij}) \bs{n}_{ij} = 0$, always holds.
Consider a perturbation $\bs{r}_i \to \bs{r}_i + d\bs{R}_i$, where $|d\bs{R}_i - d\bs{R}_j| \ll r_{ij}$.
Using the variation of the particle distance $\Delta_{ij} \equiv | \bs{r}_i + d\bs{R}_i - \bs{r}_j - d\bs{R}_j | - r_{ij}$, Eq.~\eqref{tot} can be expanded as
\eq{
    d\UU & = \sum_{i>j} \left[ \phi(r_{ij}) + \phi'(r_{ij}) \Delta_{ij} + \frac{1}{2} \phi''(r_{ij}) \Delta_{ij}^2 \right] - \UU + \order{\Delta^3} \notag \\
    & = \sum_{i>j} \left[ \phi'(r_{ij}) \Delta_{ij} + \frac{1}{2} \phi''(r_{ij}) \Delta_{ij}^2 \right] + \order{\Delta^3}. \label{UU}
}
We rewrite this series expansion in terms of the particle displacements $d\bs{R}_i$.
The perturbation of the distance $\Delta_{ij}$ is expanded as
\eq{
    \Delta_{ij} & = | \bs{r}_i + d\bs{R}_i - \bs{r}_j - d\bs{R}_j | - r_{ij} \notag \\
    & = \sqrt{ ( \bs{r}_i - \bs{r}_j + d\bs{R}_i - d\bs{R}_j )^2 } - r_{ij} \notag \\
    & = r_{ij} \sqrt{ 1 + \frac{2}{r_{ij}} \bs{n}_{ij} \cdot ( d\bs{R}_i - d\bs{R}_j ) + \frac{1}{r_{ij}^2} ( d\bs{R}_i - d\bs{R}_j )^2 } - r_{ij} \notag \\
    & = r_{ij} + \bs{n}_{ij} \cdot ( d\bs{R}_i - d\bs{R}_j ) + \frac{1}{2r_{ij}} ( d\bs{R}_i - d\bs{R}_j )^2 - \frac{1}{8} \frac{4}{r_{ij}} [ \bs{n}_{ij} \cdot ( d\bs{R}_i - d\bs{R}_j ) ]^2 - r_{ij} + \order{|d\bs{R}|^3} \notag \\
    & = \bs{n}_{ij} \cdot ( d\bs{R}_i - d\bs{R}_j ) + \frac{1}{2r_{ij}} ( d\bs{R}_i - d\bs{R}_j )^2 - \frac{1}{2r_{ij}} [ \bs{n}_{ij} \cdot ( d\bs{R}_i - d\bs{R}_j ) ]^2 + \order{|d\bs{R}|^3} \notag \\
    & = \bs{n}_{ij} \cdot d\bs{R}_{ij} + \frac{1}{2r_{ij}} (d\bs{R}_{ij}^\perp)^2 + \order{|d\bs{R}|^3}. \label{dr}
}
Substituting Eq.~\eqref{dr} into Eq.~\eqref{UU}, we obtain
\eq{
    d\UU & = \sum_{i>j} \left[ \phi'(r_{ij}) \Delta_{ij} + \frac{1}{2} \phi''(r_{ij}) \Delta_{ij}^2 \right] + \order{\Delta^3} \notag \\
    & = \sum_{i>j} \left\{ \phi'(r_{ij}) \left[ \bs{n}_{ij} \cdot d\bs{R}_{ij} + \frac{1}{2r_{ij}} (d\bs{R}_{ij}^\perp)^2 \right] + \frac{1}{2}\phi''(r_{ij}) (\bs{n}_{ij} \cdot d\bs{R}_{ij})^2 \right\} + \order{|d\bs{R}|^3} \notag \\
    & = \frac{1}{2}\sum_{i>j} \left[ \phi''(r_{ij}) (\bs{n}_{ij} \cdot d\bs{R}_{ij})^2 + \frac{\phi'(r_{ij})}{r_{ij}} (d\bs{R}_{ij}^\perp)^2 \right] + \order{|d\bs{R}|^3} \equiv \frac{1}{2}\UU_{\mr{harm}}(d\bs{R}) + \order{|d\bs{R}|^3}, 
}
where we used the force balance condition $\sum_j \phi'(r_{ij}) \bs{n}_{ij} = 0$ from the second to the third line. 
$\UU_{\mr{harm}}(d\bs{R})$ is the standard definition of the harmonic energy of a solid up to an unimportant factor of $1/2$.
To stress that $\UU_{\mr{harm}}(d\bs{R})$ is a quadratic form of $d\bs{R}$, we rewrite it as
\eq{
    \UU_{\mr{harm}}(d\bs{R}) & = \sum_{i>j} d\bs{R}_{ij}^T \left[ \phi''(r_{ij}) \bs{n}_{ij} \bs{n}_{ij}^T + \frac{\phi'(r_{ij})}{r_{ij}} \left( \id - \bs{n}_{ij} \bs{n}_{ij}^T \right) \right] d\bs{R}_{ij} \notag \\
    & = \sum_{k,l=1}^N d\bs{R}_{k}^T \left\{ \sum_{i>j} ( \delta_{ik} - \delta_{jk} ) \left[ \phi''(r_{ij}) \bs{n}_{ij} \bs{n}_{ij} + \frac{\phi'(r_{ij})}{r_{ij}} \left( \id - \bs{n}_{ij} \bs{n}_{ij}^T \right) \right] ( \delta_{il} - \delta_{jl} ) \right\} d\bs{R}_{l} \notag \\
    & \equiv d\bs{R}^T \hat{\mathcal{M}} d\bs{R},
}
where $\hat{\delta}$ is the $3\times3$ identity matrix. The $3N\times3N$ matrix $\hat{\mathcal{M}}$ is called the dynamical matrix; its eigenvalues $\lambda_\alpha$ and eigenvectors $\bs{e}_\alpha$  are of central interest in the normal mode analysis.
From the stability of the inherent structures, all the eigenvalues are positive, except the three trivial zero modes that correspond to the global translations.
In the case of a crystal, $\hat{\mathcal{M}}$ is exactly diagonalized by a discrete Fourier transform, and its eigenmodes are plane waves, which are called phonons.
However, one needs to numerically diagonalize the dynamical matrix of a amorphous solid owing to the absence of any symmetry.

The harmonic energy $\UU_{\mr{harm}}$ has two distinct contributions: the terms proportional to the second and first derivatives of the pair potential $\phi(r)$.
Their physical interpretations are as follows.
Introducing the harmonic energy, a solid is mapped to a harmonic spring network.
Then, the term proportional to $\phi''(r)$ is the contribution from an ordinary elastic energy $\propto k x^2$, where $k$ is the spring constant and $x$ is the elongation or compression of the spring.
In contrast, the term proportional to $\phi'(r)$ is the contribution from the residual force.
This is absent if all the springs are at rest when $d\bs{R}=0$.

For a given vector $\bs{e}$, $\UU_{\mr{harm}}(\bs{e})$ can further be rewritten as
\eq{
    \UU_{\mr{harm}}(\bs{e}) & = \sum_{i>j} \left[ \phi''(r_{ij}) (\bs{n}_{ij} \cdot \bs{e}_{ij})^2 + \frac{\phi'(r_{ij})}{r_{ij}} (\bs{e}_{ij}^\perp)^2 \right] \notag \\
    & = \sum_{i=1}^N \frac{1}{2} \sum_{j=1}^N \left[ \phi''(r_{ij}) (\bs{n}_{ij} \cdot \bs{e}_{ij})^2 + \frac{\phi'(r_{ij})}{r_{ij}} (\bs{e}_{ij}^\perp)^2 \right] \notag \\
    & = \sum_{i=1}^N \delta E_i = \lim_{r\to\infty} \int_{|\bs{x}|<r} d\bs{x} \sum_{i=1}^N \delta E_i \delta[ \bs{x} - (\bs{r}_i-\bs{r}^c) ] = \lim_{r\to\infty} \int_{|\bs{x}|<r} d\bs{x} \delta E(\bs{x}) = \lim_{r\to\infty} \Lambda(r).
}
Thus, the energy profile $\Lambda(r)$ naturally arises from the harmonic energy.
Strictly speaking, the location of the center $\bs{r}^c$ is arbitrary for the definition of $\delta E(\bs{x})$.
However, when $\bs{e}$ is quasi-localized, it is reasonable to appoint $\bs{r}^c$ as the core particle because it has the most negative $\delta E_i$ and typically has the strongest vibration~\cite{Shimada2018Spatial, Shimada2021Spatial}.

\clearpage

\section{Additional data for cavities}\label{sec:correlations}

\begin{figure*}[hb]
    \centering
    \includegraphics{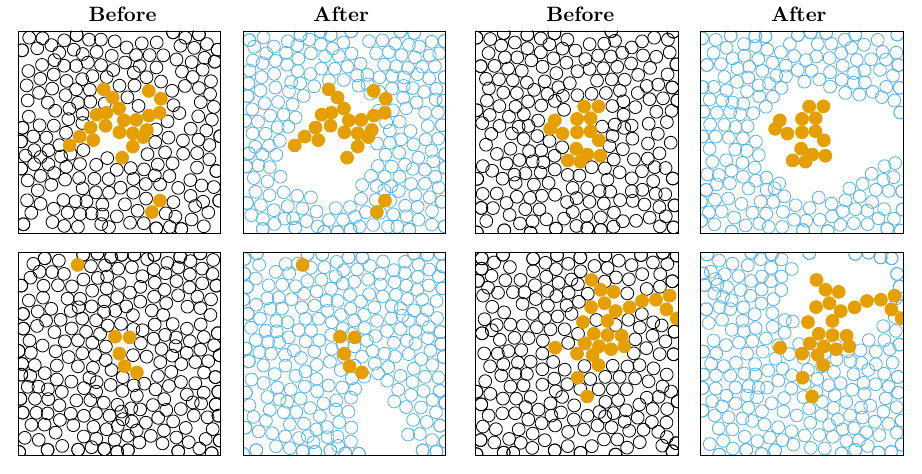}
    \caption{
    Additional configurations to supplement Fig.~\ref{fig:snap_all}.
    }
    \label{fig:app_snap_all}
\end{figure*}

In Fig.~\ref{fig:app_snap_all}, we present 2D slices of the configurations before and after the cavitation to supplement Fig.~\ref{fig:snap_all}.

\clearpage

\section{Energy profiles}\label{sec:energy}

\begin{figure*}[hb]
    \centering
    \includegraphics[width=0.9\textwidth]{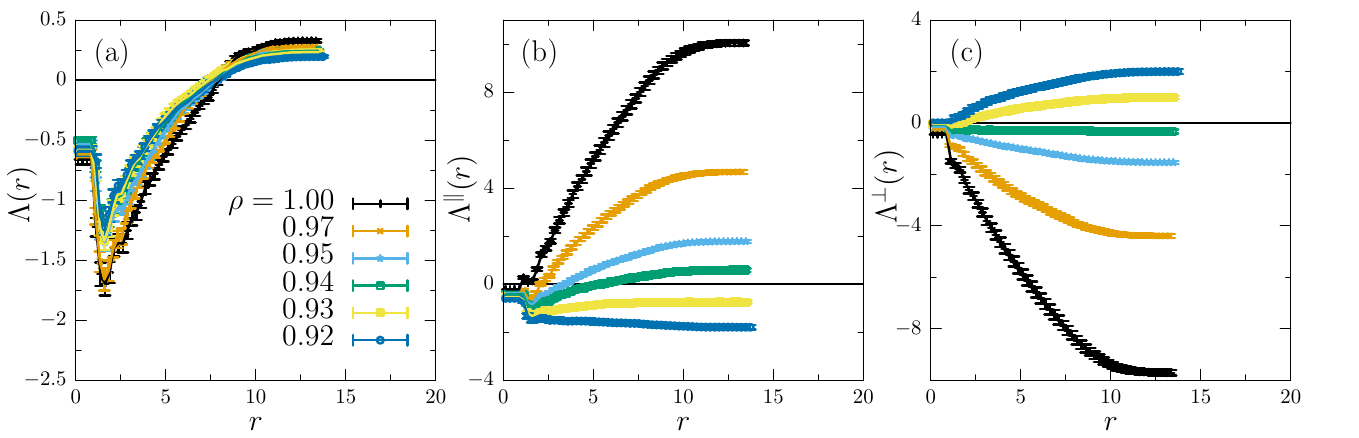}
    \caption{
    (a) Total energy profiles $\Lambda(r)$ averaged over the lowest-frequency QLM in each configuration of $N=4000$.
    (b) Energy profiles $\Lambda^\parallel(r)$ of parallel components.
    (c) Energy profiles $\Lambda^\perp(r)$ of perpendicular components.
    }
    \label{fig:app_energy_4k}
\end{figure*}

\begin{figure*}[hb]
    \centering
    \includegraphics[width=0.9\textwidth]{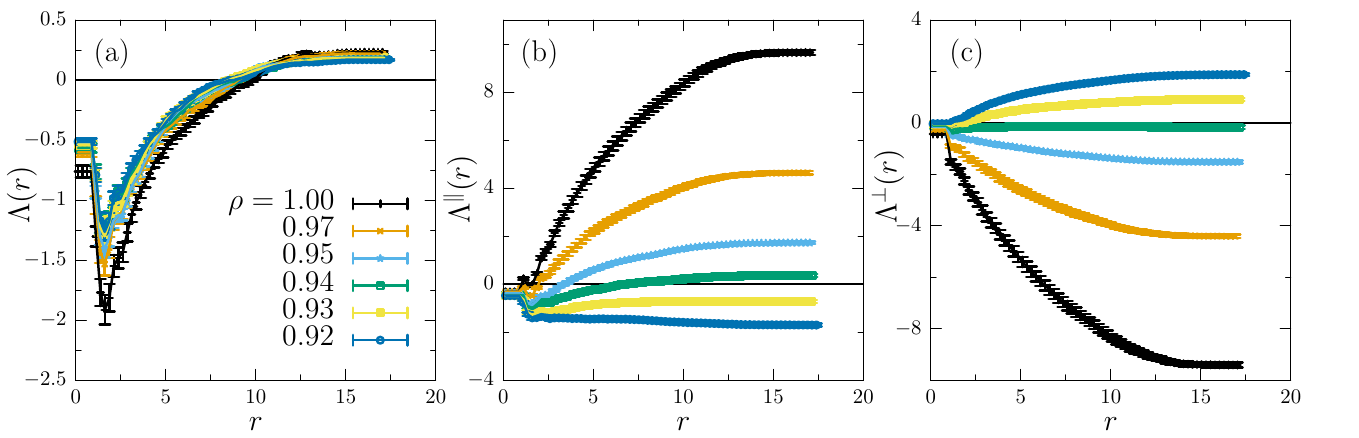}
    \caption{
    (a) Total energy profiles $\Lambda(r)$ averaged over the lowest-frequency QLM in each configuration of $N=8000$.
    (b) Energy profiles $\Lambda^\parallel(r)$ of parallel components.
    (c) Energy profiles $\Lambda^\perp(r)$ of perpendicular components.
    }
    \label{fig:app_energy_8k}
\end{figure*}

\begin{figure*}[hb]
    \centering
    \includegraphics[width=0.9\textwidth]{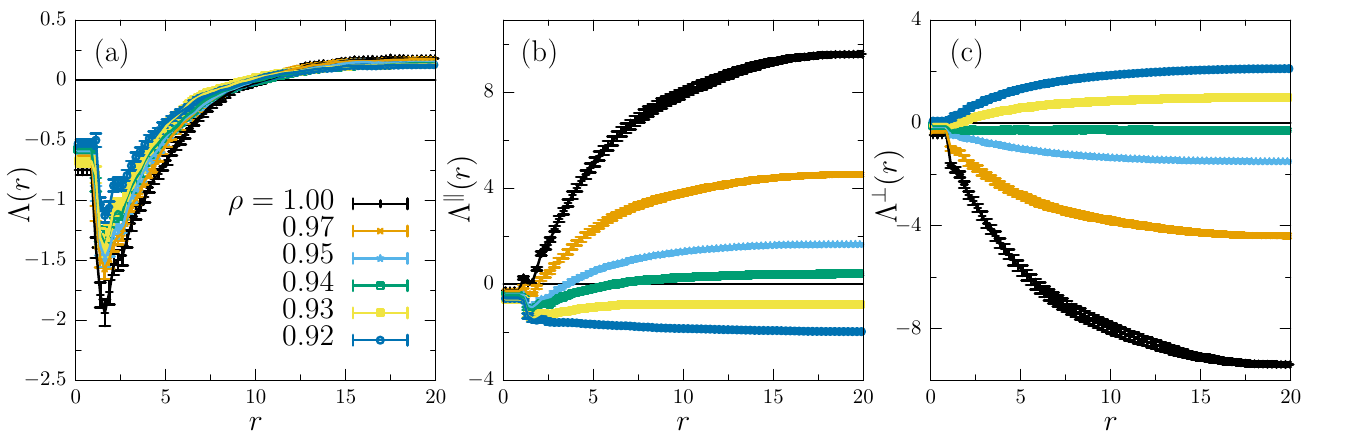}
    \caption{
    (a) Total energy profiles $\Lambda(r)$ averaged over the lowest-frequency QLM in each configuration of $N=16000$.
    (b) Energy profiles $\Lambda^\parallel(r)$ of parallel components.
    (c) Energy profiles $\Lambda^\perp(r)$ of perpendicular components.
    }
    \label{fig:app_energy_16k}
\end{figure*}

\begin{figure*}[hb]
    \centering
    \includegraphics[width=0.9\textwidth]{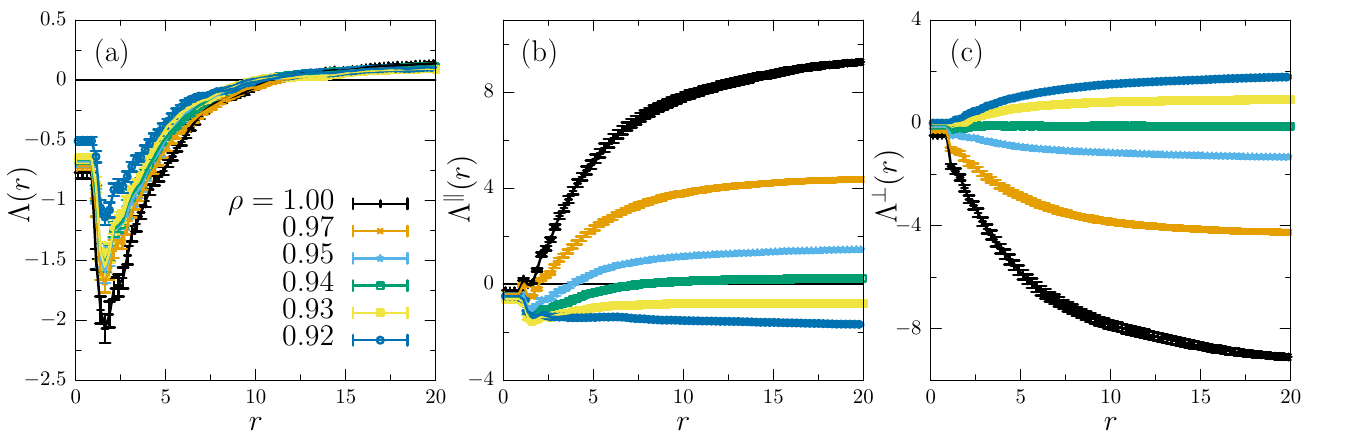}
    \caption{
    (a) Total energy profiles $\Lambda(r)$ averaged over the lowest-frequency QLM in each configuration of $N=32000$.
    (b) Energy profiles $\Lambda^\parallel(r)$ of parallel components.
    (c) Energy profiles $\Lambda^\perp(r)$ of perpendicular components.
    }
    \label{fig:app_energy_32k}
\end{figure*}

In this section, we present the energy profiles of the systems for different $N$ and $\rho$. 
In Figs.~\ref{fig:app_energy_4k}, \ref{fig:app_energy_8k}, \ref{fig:app_energy_16k}, and \ref{fig:app_energy_32k}, we compare the energy profiles of $N=4000$, 8000, 16000, and 32000, respectively.
Different symbols represent different densities as shown in the legend.
We do not observe any qualitative difference among these system sizes.

\clearpage
\twocolumngrid

%

\end{document}